\documentclass[pdflatex,sn-mathphys-num,referee]{sn-jnl}% Math and Physical Sciences 

\usepackage{graphicx}%
\usepackage{multirow}%
\usepackage{amsmath,amssymb,amsfonts}%
\usepackage{amsthm}%
\usepackage{mathrsfs}%
\usepackage[title]{appendix}%
\usepackage{xcolor}%
\usepackage{textcomp}%
\usepackage{manyfoot}%
\usepackage{booktabs}%
\usepackage{algorithm}%
\usepackage{algorithmicx}%
\usepackage{algpseudocode}%
\usepackage{listings}%
\usepackage{graphicx}
\usepackage{tikz}
\usepackage{pgfplots}
\usepackage{comment}
\usepackage{lineno}
\usepackage{cleveref}
\usepackage{float}
\usepackage{stfloats}
\graphicspath{ {./figs/} }

\usepackage{draftwatermark}
\SetWatermarkText{Under Review}
\SetWatermarkScale{4}          % Larger text

\theoremstyle{thmstyleone}%
\theoremstyle{thmstyletwo}%

\theoremstyle{thmstylethree}%

\begin{document}

\title[Article Title]{Demonstration of a mechanical external biventricular assist device for resuscitative thoracotomy}

\author*[1]{\fnm{Kristóf} \sur{Sárosi}}\email{ksarosi@ethz.ch}

\author[1]{\fnm{Thomas} \sur{Kummer}}

\author[1]{\fnm{Thomas} \sur{Roesgen}}

\author[2,3]{\fnm{Stijn} \sur{Vandenberghe}}

\author[3,4]{\fnm{Stefanos} \sur{Demertzis}}

\author[1]{\fnm{Patrick} \sur{Jenny}}

\affil*[1]{\orgdiv{Institute of Fluid Dynamics, Department of Mechanical and Process Engineering}, \orgname{ETH Zurich}, \orgaddress{\city{Zurich}, \country{Switzerland}}}

\affil[2]{\orgdiv{Cardiovascular Engineering, Cardiocentro Ticino Institute}, \orgname{Ente Ospedaliero Cantonale}, \orgaddress{\city{Lugano}, \country{Switzerland}}}

\affil[3]{\orgdiv{Faculty of Biomedical Sciences}, \orgname{Università della Svizzera italiana},  \orgaddress{\city{Lugano}, \country{Switzerland}}}

\affil[4]{\orgdiv{Cardiac Surgery \& Cardiovascular Engineering, Cardiocentro Ticino Institute}, \orgname{Ente Ospedaliero Cantonale},  \orgaddress{\city{Lugano}, \country{Switzerland}}}

%%==================================%%
%%             Abstract             %%
%%==================================%%

\abstract{\textbf{Purpose:} Resuscitative thoracotomy, a high-risk procedure involving open heart massage, serves as a last resort for life-threatening conditions like penetrating chest wounds, severe blunt trauma, or surgery-related cardiac arrest. However, its success rate remains low, even when primarily carried out by highly trained specialists. This research investigates the potential of an external biventricular assist device (BiVAD). By replacing open heart massage with our BiVAD device during resuscitative thoracotomy, we aim to achieve sufficient cardiac output, maintain physiological pressure levels, and potentially improve patient survival in these critical situations.

\textbf{Methods:} The proposed BiVAD system features a simple 3D printed patch design for direct cardiac attachment, an actuation device, and a vacuum pump. The straightforward design allows quick application in emergency situations. The BiVAD system was tested in a hydraulic mock circulation, utilizing a silicone heart as ex vivo porcine heart measurements were deemed inconclusive. Three actuation modes were tested for proof-of-concept: manual patch actuation, standard cardiac hand massage, and utilizing full capabilities of our BiVAD patch system with actuation device operation. Overall performance was assessed on ventricular pressure and flow rate data.

\textbf{Results:} Focusing on achieving the optimal cardiac output of 1.5 L/min (critical for patient survival), we tested our patch system against cardiac hand massage at a fixed rate of 60 bpm. The results include raw and statistically evaluated flow rate and pressure measurements for both the left and the right ventricle. Notably, our BiVAD system not only achieved to operate in the range of required cardiac output but also significantly reduced peak pressure in both ventricles compared to standard cardiac hand massage.

\textbf{Conclusion:} This initial evaluation using a silicone heart model demonstrates the potential of our BiVAD system to achieve sufficient cardiac output while reducing peak pressure compared to cardiac hand massage. Further development holds promise for effective cardiac support in resuscitative thoracotomy.
}


\maketitle

%\linenumbers
%%==================================%%
%%           Introduction           %%
%%==================================%%

\section{Introduction}\label{intro}

%Cardiovascular diseases, resc thoracotomy
Cardiovascular diseases remain the leading cause of mortality in the developed world. Despite significant technological advancements, the proportion of affected individuals within our population continues to rise \cite{Callan2020}. These diseases present a diverse spectrum of challenges, requiring tailored medical and engineering solutions. In extreme life-threatening circumstances, resuscitative thoracotomy is an emergency procedure used to address potential causes of cardiac arrest, such as cardiac tamponade, intrathoracic bleeding, and the need for open heart massage \cite{Weare2024}. While historically the primary technique for managing heart failure, open heart massage is now used only in specific cases like penetrating chest trauma, pericardial tamponade, or cardiac arrest following chest surgery \cite{Souza}. Despite requiring a trained surgeon, the survival rate of open heart massage remains low \cite{Dayama2016}, and the procedure ties up crucial medical personnel \cite{Weare2024}. Due to device availability, the number of patients who can live with heart failure has increased in the last decades \cite{Callan2020}. As the number of patients with heart failure is increasing in developed countries \cite{Callan2020}, devices like ventricular assist devices (VADs) are essential for the survival of the patients.

%devices
%VADs history
One of the initial goals of VAD design was to replace open heart massage with a machine capable of maintaining long-term blood flow. In the 1950s and 60s, this led to the concept of direct mechanical cardiac assistance \cite{Vineberg1957} \cite{McCabe1983} \cite{Lowe1991} \cite{Anstadt1991}. Later, closed chest CPR replaced open heart massage for most cases, and modern technology enabled implantable, small-scale turbo machines to deliver blood flow. While highly effective in bridge-to-transplant scenarios \cite{Eisen}, these devices have a significant drawback: they require invasive heart surgery and operate in direct contact with the patient's blood \cite{Oz2002} \cite{Melvin1996} \cite{Long2019}. This increases the risk of complications, limiting their suitability for many patients. Furthermore, most devices support only the left ventricle (LVADs), causing pressure imbalances within the heart that can lead to aortic insufficiency \cite{Kagawa2020}, right heart failure \cite{Hall2022}, and septal shift \cite{Guglin2020}. Direct contact with the blood not only can lead to infections \cite{Tattevin2019}, but also requires medication to tackle hemorrhagic and thromboembolic complications \cite{Lowe1991}, further restricting patient eligibility. Available devices require invasive techniques, special medication, and target bridge-to-transplant and long-term solutions. This leaves many without a safe short or mid-term solution. Callan \cite{Callan2020} provides a comprehensive overview of available devices.

%current research
The development of novel approaches for blood-contacting and non-blood-contacting devices are an active area of research. Despite advancements in VAD technology, challenges remain in optimizing blood flow while minimizing device-related complications. Research directions include a pulsating pump using a hemocompatible membrane \cite{Ferrari2021}, an invasive soft robotic device moving the ventricular walls \cite{Payne2017Science}, further improvements of small-scale turbo machines \cite{Varshney2022}, and new approaches in direct mechanical ventricular actuation \cite{Hord2023}. Despite the lack of success in direct mechanical ventricular actuation, many variants are under development. Interestingly, most of them were built on the concept of the Anstadt cup \cite{Anstadt1991} \cite{Jagschies2018} \cite{Letsou2022}. A potential challenge with these devices lies in precisely controlling pressure within both ventricles. Non-physiological pressure levels or imbalances between the left ventricle (LV) and right ventricle (RV) can lead to complications. The RV is particularly vulnerable to excessive pressure due to equal force application on both ventricles by direct compression. In previous studies, we also noticed that the compression of the LV is delayed until the RV is compressed completely. However, from publications on direct mechanical ventricular actuation devices, we lack an understanding of these phenomena and their effect on the cardiovascular system. Most reports only focus on aortic pressure and cardiac output, limiting available data on both ventricles. This makes comprehensive evaluation of these devices difficult. 

\begin{figure}[h]%
\centering
\includegraphics[width=0.5\columnwidth]{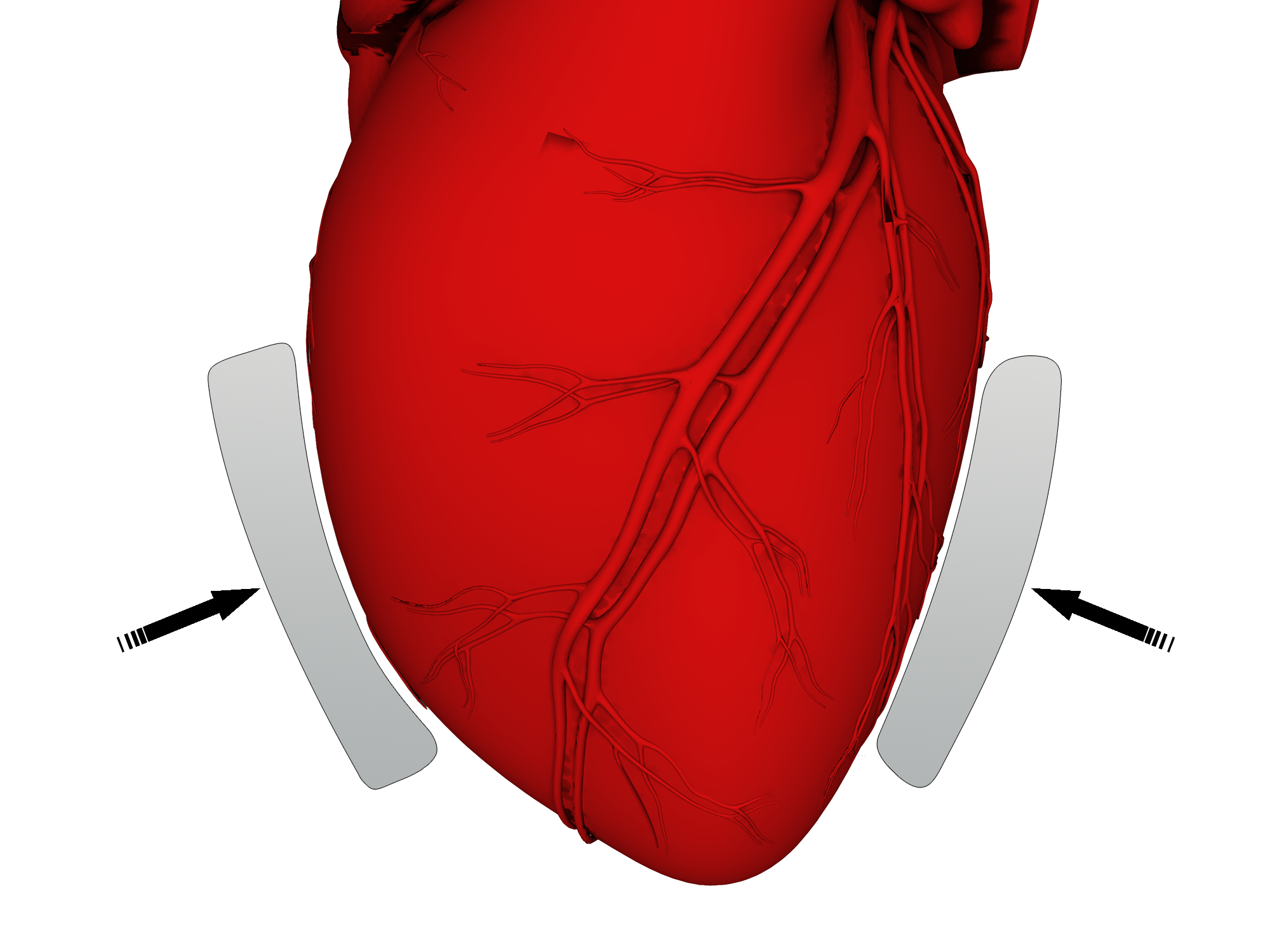}
\caption{Direct mechanical ventricular actuation using an epicardial patch system. The patches compress the ventricles by exerting force on the epicardium.}
\label{fig:device_schematics}
\end{figure}

%our device
This work introduces an early prototype of a short-term emergency biventricular assist device (BiVAD) designed to replace open heart massage with direct mechanical ventricular actuation. Although the original Anstadt cup was meant to replace open heart massage, the aforementioned direct mechanical ventricular actuation devices aim to provide long-term solutions. While using a similar actuation concept, our device addresses the unmet need for rapid intervention in resuscitative thoracotomy. Direct mechanical ventricular actuation requires no blood-thinning medication as the device is not in direct contact with the patient's blood. As the success rate of open heart massage in resuscitative thoracotomy is currently low, we aim to improve on that by introducing little to no new complication factors. The goal of our research is to develop a BiVAD concept that simultaneously maintains physiological pressure levels for both ventricles while ensuring sufficient cardiac output. Here we present a 3D-printed epicardial patch system actuated by servo motors via wire cables. \Cref{fig:device_schematics} demonstrates the essential idea of direct mechanical ventricular actuation with our patch system. The design allows quick application in emergency cases, freeing medical staff from open heart massage. Additionally, thanks to its straightforward design, the BiVAD requires minor training for application or operation. This BiVAD system aims to offer control over pressure and flow rates, facilitating the exploration of optimal resuscitation strategies.

%experiments
To experimentally validate the BiVAD concept, we build upon the simulation framework by Kummer et al. \cite{Kummer2022}, which explored the interaction of a failing heart's digital twin with a BiVAD patch system. The potential of a BiVAD patch system was already demonstrated in their study, albeit only in a simulation framework. Their lumped parameter model of human circulation is replicated on a test bench using components such as ball valves, compliance chambers, and check valves. Pressure and flow rates were measured simultaneously for both ventricles. Comparative studies of open heart massage and mechanical actuation with the patch system were conducted on silicone hearts. Initial experiments with porcine heart samples were inconclusive, leading to the choice of silicone models. The digital twin of Kummer et al. \cite{Kummer2022}, the silicone heart, and the geometry of the patches were based on the same generic human heart CAD model \cite{3d}. This proof-of-concept study aimed to demonstrate the BiVAD patch system as a viable alternative to open heart massage.

%results
This study demonstrated the feasibility of our BiVAD system using a silicone heart model. At a fixed rate of 60 bpm, we compared our system against cardiac hand massage, focusing on the critical 1.5 L/min cardiac output target. Results include raw and statistically evaluated pressure and flow data for both ventricles. Importantly, our BiVAD achieved to operate in the range of the target cardiac output while significantly reducing peak ventricular pressures compared to cardiac hand massage. Additionally, the BiVAD system reliably maintained the desired actuation rate, cardiac output and peak pressure performance.

%summary
The paper is organized as follows: Section \ref{methods} introduces the BiVAD system (including patches, actuation, and vacuum pump), the test bench setup, data collection methods, experimental design, and data analysis. Section \ref{results} presents the experimental results, including raw and statistically evaluated data. Section \ref{discussion} analyzes the findings, compares BiVAD performance to manual methods, and identifies system strengths and weaknesses. Section \ref{limit} acknowledges experimental limitations and outlines future research directions. Section \ref{conclusions} summarizes the study's key findings and their implications.

%%==================================%%
%%             Methods              %%
%%==================================%%

\section{Methods}\label{methods}

%general introduction of the concept
This section provides a detailed breakdown of the BiVAD system, the experimental setup, and the methodology. 

Our goal was to design a BiVAD that provides sufficient blood flow in the circulation while keeping pressure levels safe in both ventricles without damaging the epicardium. When imposing external actuation through the epicardium, it is important that not only the flow rates but also the pressures remain in sustainable ranges throughout the actuation, while keeping the coronary and capillary flow unobstructed. In extreme cases, the heart can deform in such a way that the tissue or the blood vessels get damaged. The deformation may cause severe valve leakage or even rupture the chordae tendinae. We used the initial design of Kummer et al. and modified it to address the issues they noticed in their studies.

\begin{figure*}[h]%
\centering
\includegraphics[width=0.5\columnwidth]{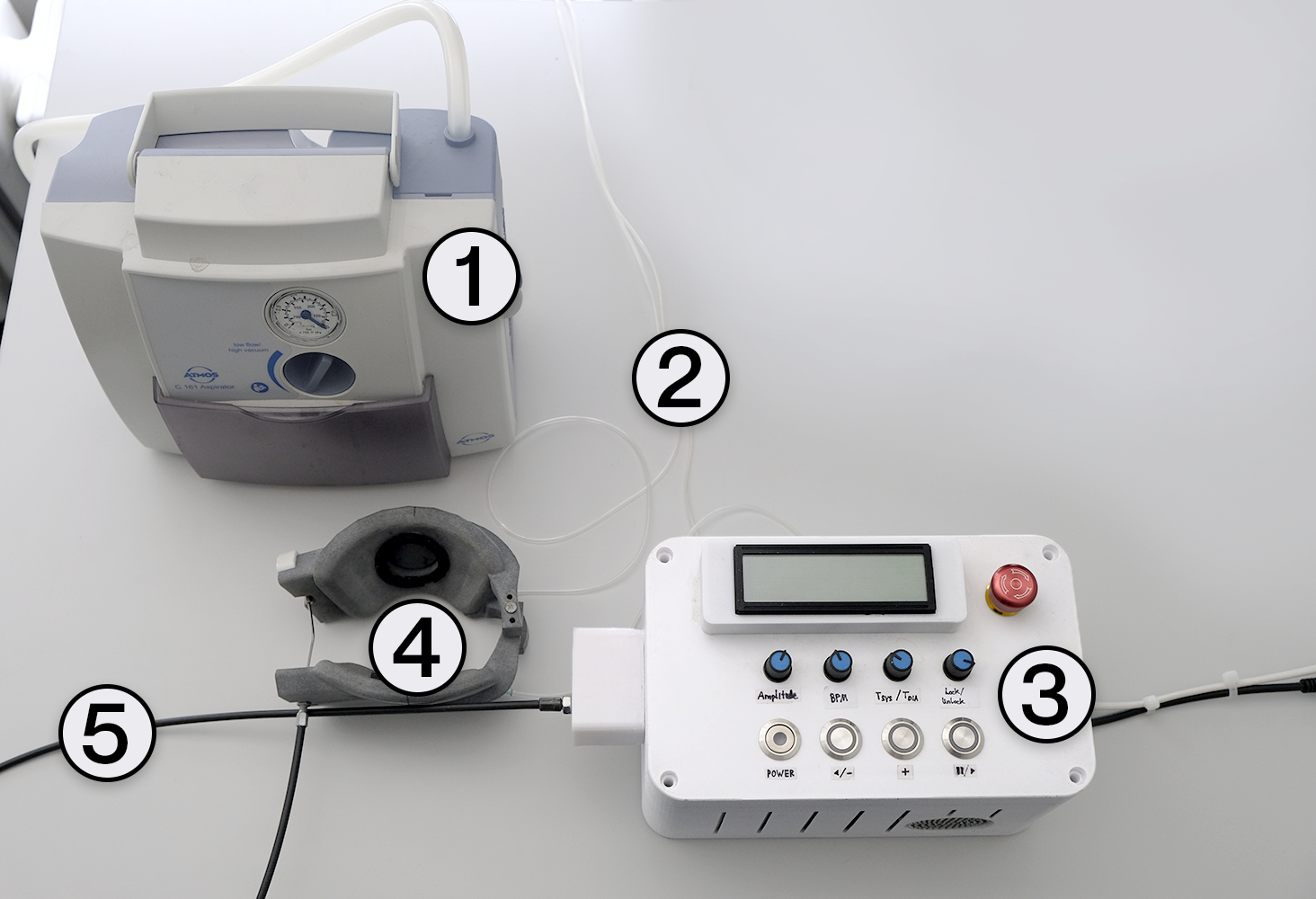}
\caption{The system, including the vacuum pump (1), the actuation device (3), and the patches (4). The patches are in direct contact with the heart, and the actuation is done by two servo motors connected to a single Bowden cable (5). The distance between the patches is adjustable to the size of the heart, and the actuation also is variable in terms of amplitude, frequency, and the ratio of systolic and diastolic duration. The vacuum pump is connected to the outer surface of the patches via vacuum lines (2), where the connection points of internal vacuum channels are located.}
\label{fig:package}
\end{figure*}

The BiVAD system includes patches that directly compress the heart, an actuation device to control the applied force, and a vacuum pump to ensure the patches remain securely attached to the epicardium. This system is presented in \cref{fig:package}. The patches are in direct contact with the heart and are responsible for compressing the heart and controlling the compression in different regions. The actuation device provides the total compression force for the patches which can be adjusted to remain in the desired range. The vacuum pump is connected to the patches, which each have a vacuum channel to create suction inside the central circular cavity of the patches. Without this suction, the devices can easily slip off the wet epicardium.

%patches
\subsection{Patches}\label{patches}

%design considerations  
Using the device in resuscitative thoracotomy cases implies different requirements than for controlled chronic implantation. As the patient would be in a life-threatening state, the device must be easy to apply in a matter of minutes. Our design can be applied quickly without special tools, and the locking mechanism is robust enough to withhold the pulling force of the device. Providing sufficient blood flow while keeping the pressure in a healthy range is the most crucial and hardest goal to achieve for such a device. Therefore, our ambition is to show that the patches are capable of handling this issue.

\begin{figure}[h]%
\centering
\includegraphics[width=0.45\columnwidth]{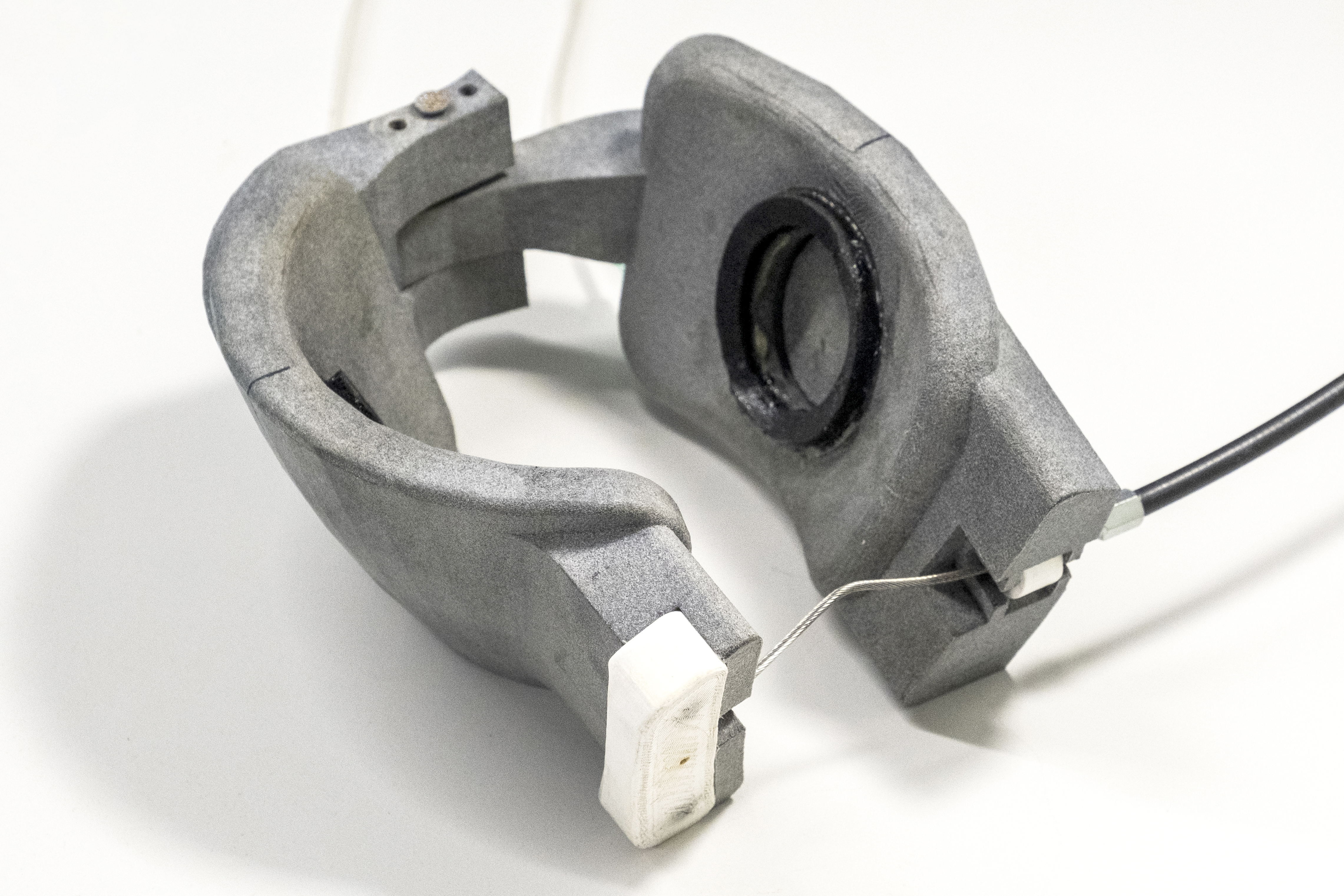}
\includegraphics[width=0.45\columnwidth]{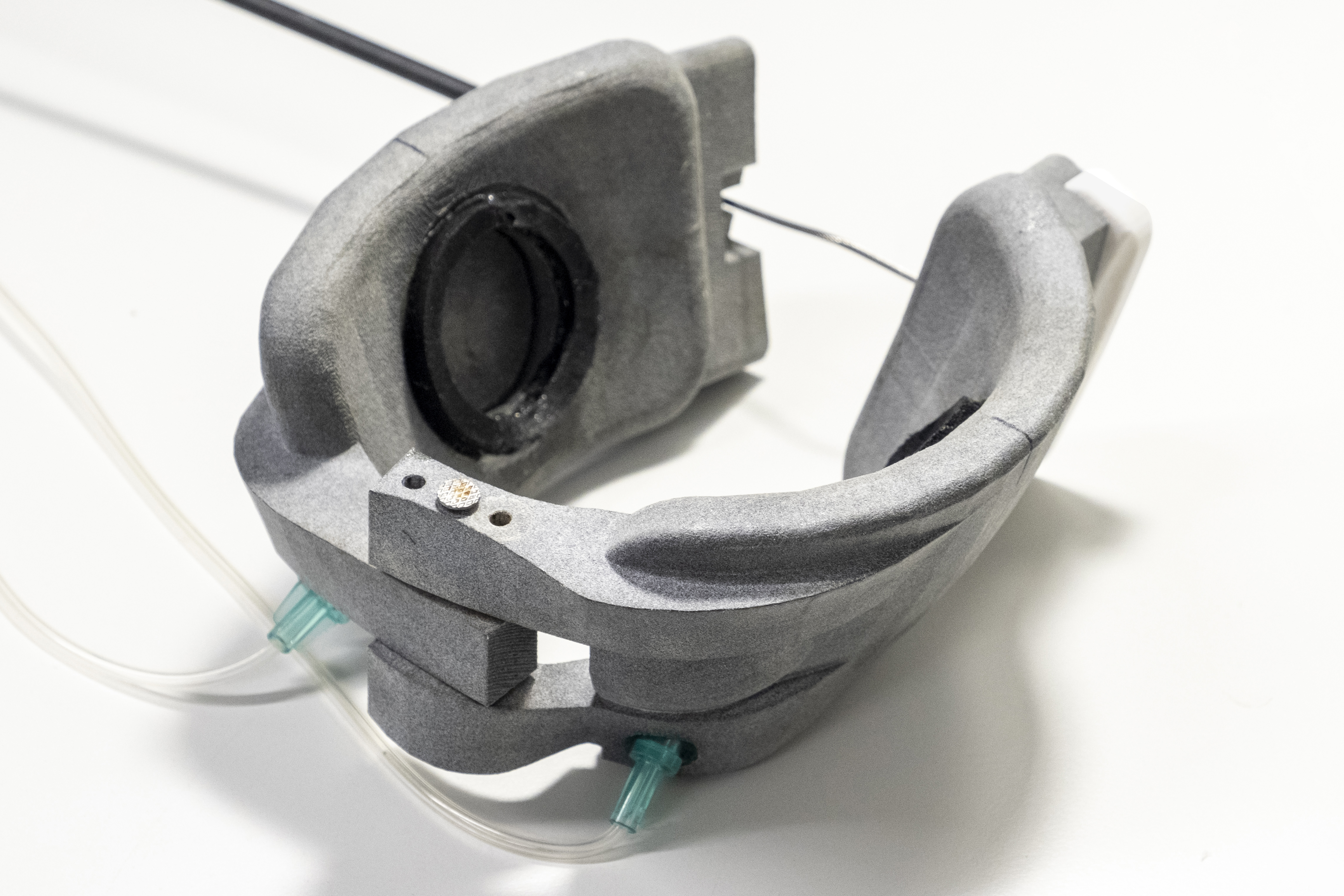}
\caption{Two views of the current patch system. The front view shows the magnetic clamp system of the Bowden cable, and the back view shows the connections to the vacuum pump. The shape was defined using a 3D heart model \cite{3d}. The patches are 3D printed, and the distance between the two halves is adjustable by a pin. The axis between the two patches was implemented to avoid the rotation of the patches, thus the ballooning of the apex. The extension on the side of the patches, where the Bowden cable connects to the patches, was necessary to keep the wire away from the surface of the heart. The vacuum pump creates a low-pressure area inside the sealing rings of the patches through the vacuum lines. This is essential to keep the patches on the heart.}
\label{fig:patches}
\end{figure}

%actual device design 
The latest patch design can be seen in \cref{fig:patches}. The shape of the patches is based on the surface of a general 3D heart model \cite{3d}. The same 3D model was used to produce the silicone heart samples. On the one hand, it is important to note that this design retains its specific form and does not conform to individual patient heart shapes. On the other hand, the heart consists of elastic tissue and can create better contact with the patches as it deforms. Some specific design choices are worth mentioning. The axis connecting the two sides limits the DoF of the compressing motion. With this, the patches cannot tilt, a phenomenon that led to the ballooning of the apex using previous prototypes. The extension, where the actuator’s Bowden cable connects to the device, helps keep the wire away from the surface of the heart. The wire cable could cut into the heart tissue without the extension. The patches include a vacuum channel, leading to the cavity in the centre of the patches. Inside the sealing ring, a low-pressure area is created to keep the patches on the surface of the heart.

The patches were 3D printed, using HP PA12 material in a Jet Fusion process. As the patches are not shape-conforming, the axis has multiple holes to adjust the distance between the two halves. As an early prototype, the biocompatibility of the material was not of concern at the current stage.

%actuation
\subsection{Actuation}\label{actuation}

The actuator includes two digital servo motors. These motors are connected to a single Bowden cable transmitting the force to the patches. The two servo motors are controlled by two Arduino devices which adjust amplitude, frequency, and the systolic and diastolic periods of the actuation.

The device can be controlled by knobs or via WiFi, which provides a lot of flexibility and helps to find the optimal parameter combination.

The actuation is not the main focus of this research, but it is essential for the performance of the whole system. As the cardiac output strongly correlates with the power output of the motors of the actuation system, improving its output is important, but rather technical.

More details about the actuation components can be found in Appendix \ref{A_actuation}.

%vacuum pump
\subsection{Vacuum Pump}\label{pump}

The vacuum pump is an ATMOS C161 aspirator, capable of providing 750 mmHg negative differential pressure to keep the patches attached to the heart. The vacuum line is directly connected to the patches, with internal vacuum lines leading to the circular cavities in the centre of the patches.

%test bench
\subsection{Test Bench}\label{testbench}

\begin{figure}[h]%
\centering
\includegraphics[width=0.5\columnwidth]{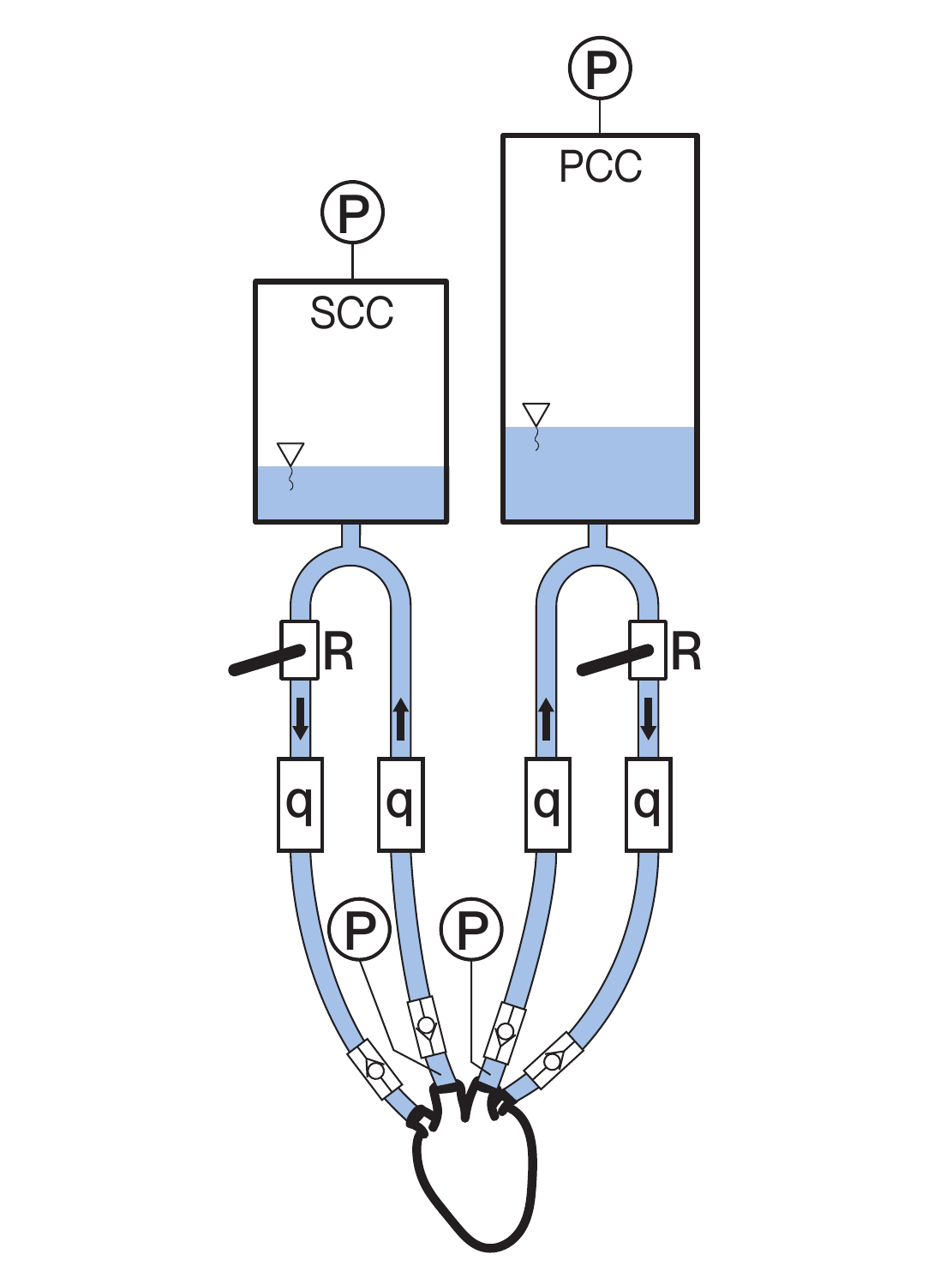}
\caption{The schematics of the test bench. \textit{SCC} stands for Systemic Compliance Chamber, \textit{PCC} for Pulmonary Compliance Chamber, \textit{P} for static pressure taps, \textit{q} for ultrasonic flow meters, and \textit{R} for adjustable resistances.}\label{fig:testbench}
\end{figure}

This study utilizes a hydraulic analogue of human circulation for testing cardiac assist device performance. \Cref{fig:testbench} presents the setup, highlighting the most important parts. The test bench consists of a systemic and a pulmonary circuit, a water pump, and the sensors. Each circulation has a compliance chamber modeling the compliance of the blood vessels, and a ball valve modeling the flow resistance of the system. Compliances and resistances are adjustable. Distilled water serves as a blood analogue. 

Accurate data acquisition is paramount in medical device testing. The system employs ultrasonic flow meters to measure rapid flow changes without altering the setup.  Analog pressure sensors record data from the left and right ventricles, and compliance chambers. The analog pressure sensors are connected to the left ventricle, the right ventricle, and the two compliance chambers. The four ultrasonic flow sensors are included in the circulation, one in each main blood vessel of the heart. Two Arduino Portenta H7 boards simultaneously capture flow rates and pressures at 100Hz / ADC 16bit.

For further information on components, see Appendix \ref{A_data}.

%silicone hearts
\subsection{Silicone Heart}\label{hearts}

\begin{figure}[h]%
\centering
\includegraphics[width=0.5\columnwidth]{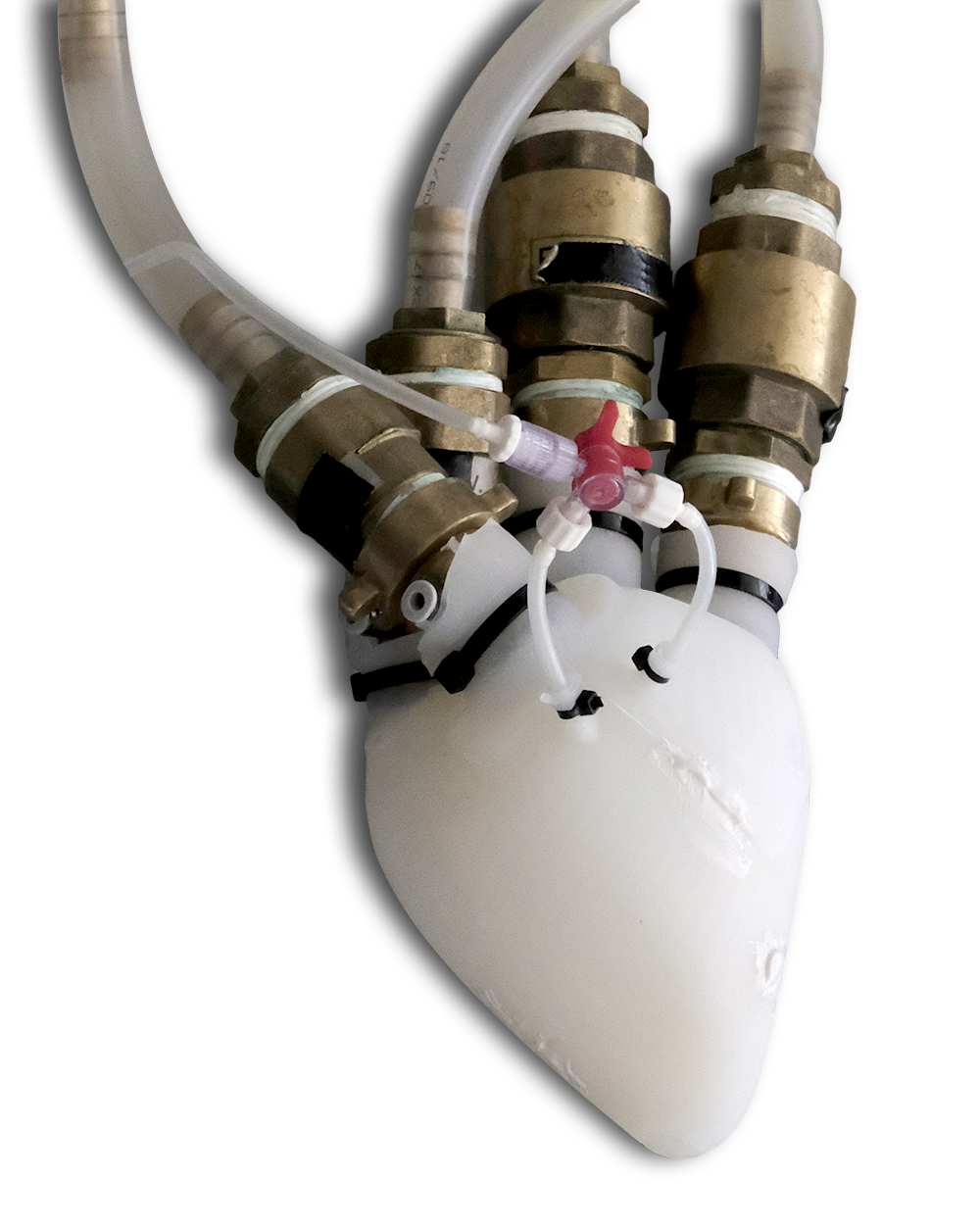}
\caption{The silicone heart used in the experiments. The unit was cast using \textit{Dragon Skin \texttrademark} silicone. The pressure taps can be found on the brass connectors. The check valves differ for different blood vessels.}
\label{fig:hearts}
\end{figure}

For the silicone heart, we used the material \textit{Dragon Skin \texttrademark}. This silicone provides a soft material while providing flexibility and durability for our pressure ranges and our applications. The silicone hearts are cast and closed using silicone adhesive. The silicone heart used in the experiments can be seen in \cref{fig:hearts}.

In Section \ref{results}, we discuss results using silicone heart samples.

%experimental design
\subsection{Experiments}\label{experiments}

\begin{figure*}[htbp!]%
\centering
\includegraphics[width=0.3\textwidth]{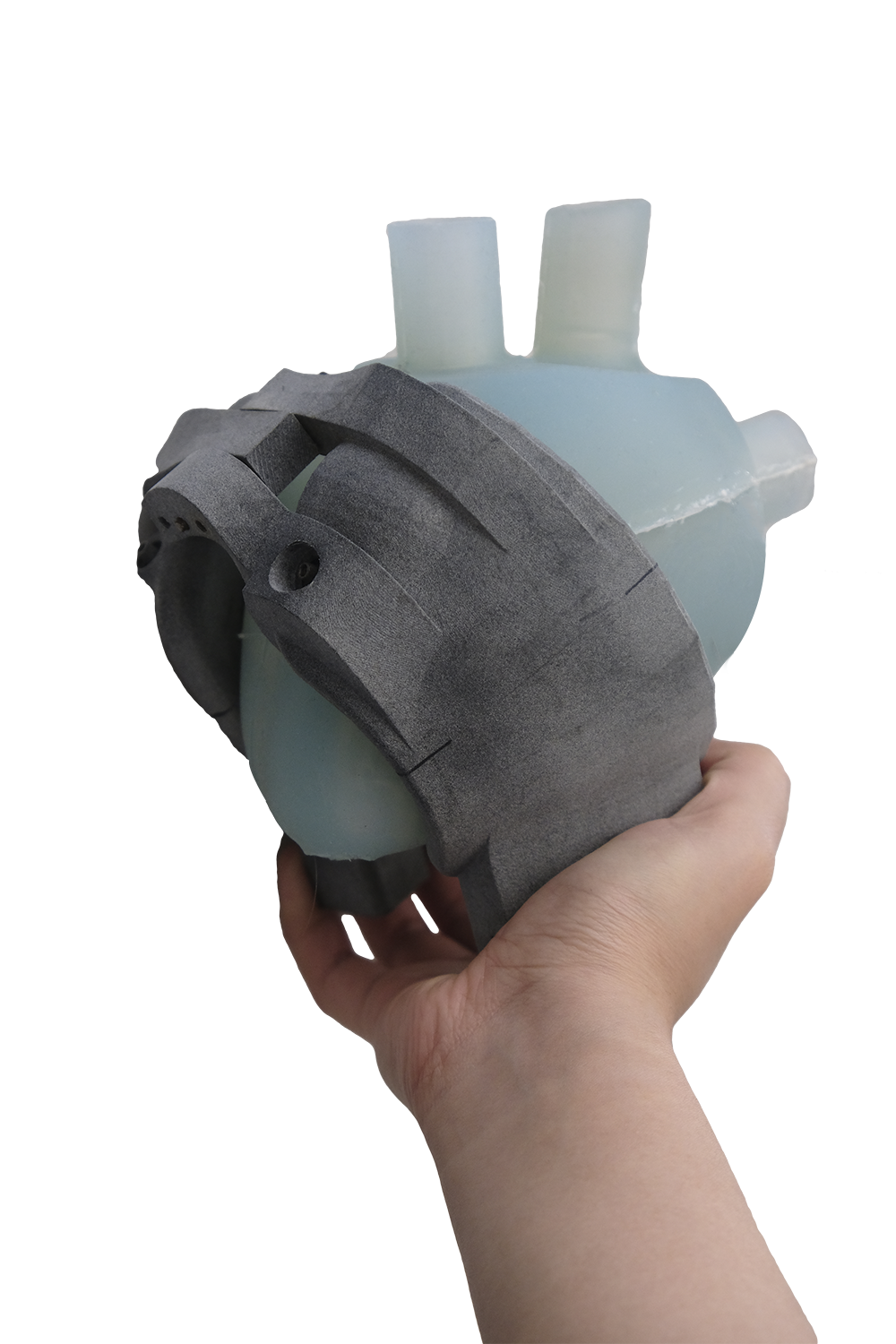}
\includegraphics[width=0.3\textwidth]{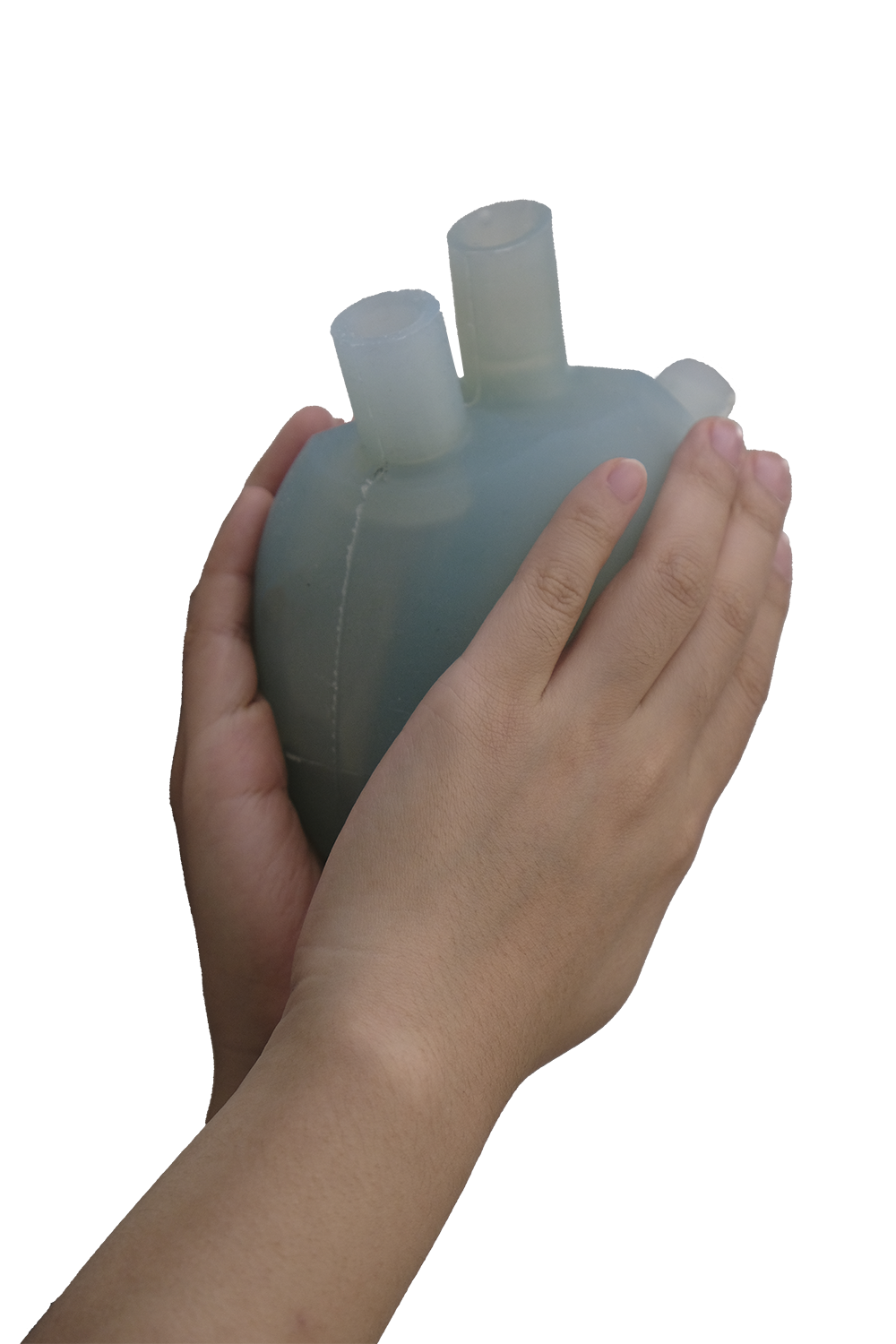}
\includegraphics[width=0.3\textwidth]{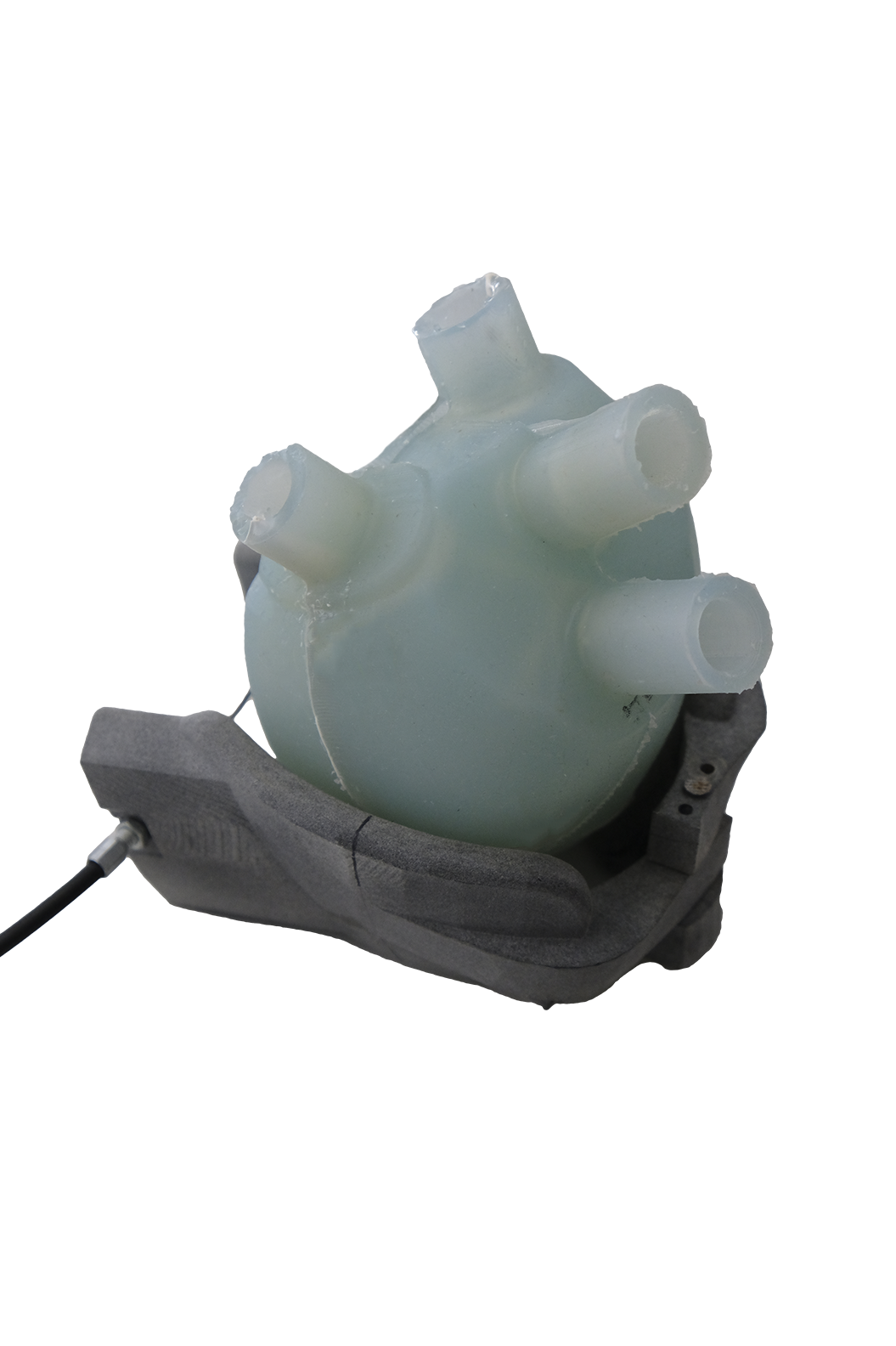}
\caption{The three modes of actuation. \textbf{Left}: manual compression of the patches, \textbf{Centre}: cardiac hand massage,  \textbf{Right}: compression of the patches by the actuation device}
\label{fig:modes}
\end{figure*}

This study compared three actuation modes: manual patch compression, cardiac hand massage (CHM), and actuation via a specialized device (\cref{fig:modes}). Manual patch compression demonstrates the system's capabilities without the actuation device. Amplitude, frequency, and $T_{systolic}/T_{diastolic}$ are adjustable using the actuation device. In both manual patch compression and CHM, the only variable is the rate of actuation. Given the use of a silicone heart model, synchronization with a heartbeat was not required, also cardiac output and pressure are supplied entirely by the mode of actuation. During patch device actuation, the patches were secured to the heart's surface with the locking mechanism, and the vacuum pump was engaged.

\subsection{Data-Processing}\label{postpro}

To ensure robust statistical analysis, collected data was segmented into individual heart cycles. Each presented measurement lasted one minute, yielding over 50 heart cycles per configuration. To evaluate the expected performance  (see \cref{fig:silicone_mean}) of the three modes, the individual cycles were resampled to their median length to neglect the inconsistency of the rate of actuation (see \cref{fig:silicone_box_pulse}). Understanding the actuation mode's influence on actuation frequency, cardiac output, and peak pressure variability is crucial. Due to differences in consistency between the three modes, the data did not meet homoscedasticity assumptions, precluding the use of ANOVA. Therefore, Welch's t-test was used to compare expected values, and Levene's test was employed to assess variance. Box plots displaying significance use the symbols  $\mu$ and $\sigma$  (notations outlined in Table \ref{table:A_signif}, Appendix \ref{A_stats}).

%%==================================%%
%%             Results              %%
%%==================================%%

\section{Results}\label{results}

In demonstrating the performance of a VAD, the most important factors are cardiac output, pressure values, damage to the tissue, and reliability. The selected results have matching frequency and orientation of the device to provide a fair comparison for all three modes of actuation (see \cref{fig:modes}). All of the following cases were recorded at 60 bpm. 

The following values were recorded during the experiments: flow rate (aorta, pulmonary artery, pulmonary vein, vena cava) and pressure (aorta, pulmonary artery). Regarding cardiac output, we found that actuating the heart at 60 bpm was the ideal rate of actuation. The orientation of the device is such that the heart sits perfectly between the patches (the intended orientation during the design phase). Important to note that the results are shown for non-self-actuating silicone hearts, meaning that the flow and pressure output are caused entirely by the mode of actuation.

\begin{figure*}[htbp!]
    \centering
    \includegraphics[scale=1]{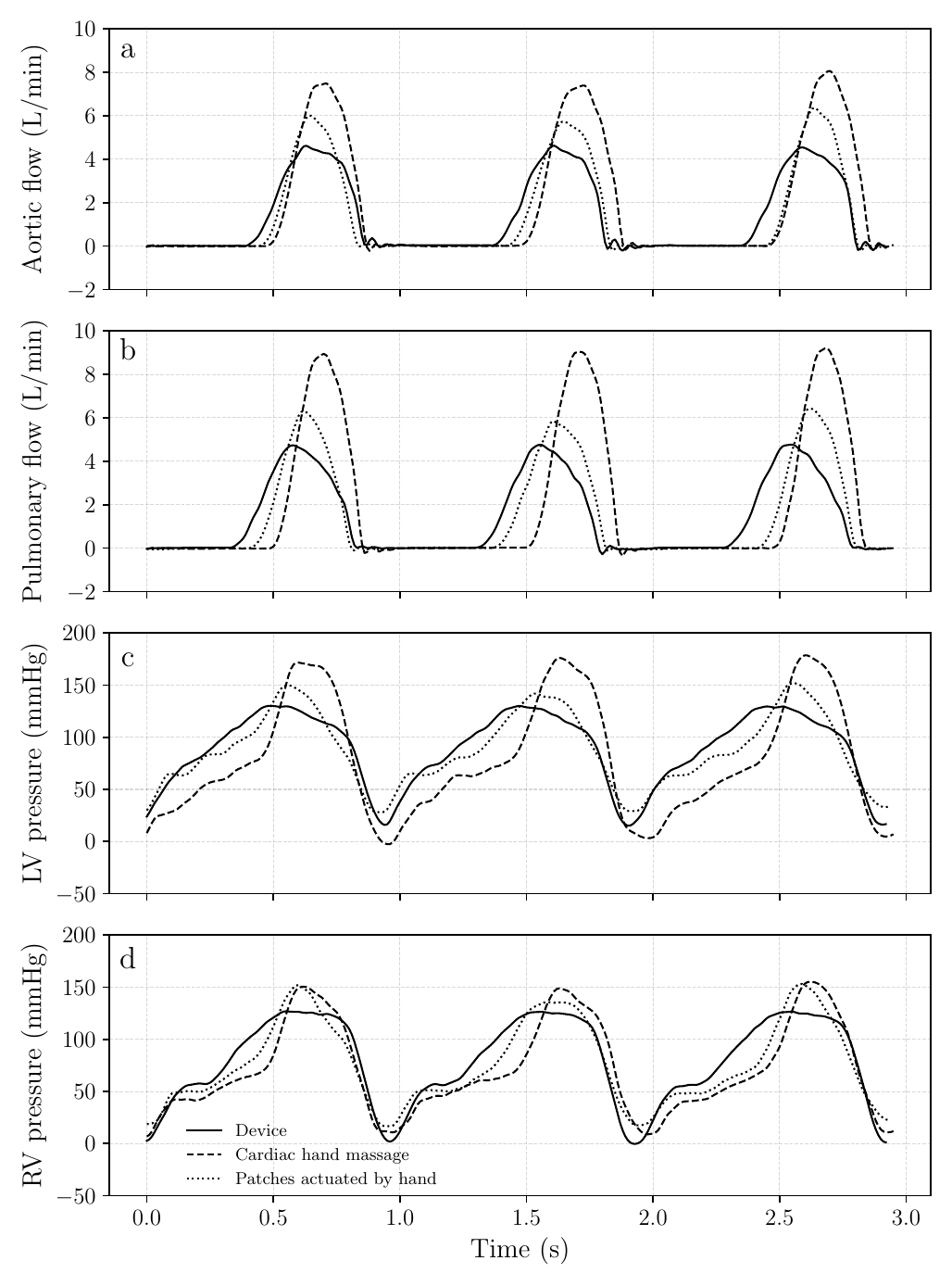}
    \caption{Raw data of the three actuating modes (see \cref{fig:modes}) for three heart cycles. For line style, please refer to \textbf{d}. \textbf{a}: flow rates in the aorta. \textbf{b}: flow rates in the pulmonary artery. \textbf{c}: pressures in the left ventricle. \textbf{d}: pressures in the right ventricle. The cardiac hand massage produces both higher pressure and flow rate than the other two modes. Actuating the patches by hand also produces a higher flow rate and pressure than using the actuation device.}
    \label{fig:silicone_raw}
\end{figure*}

For a proof-of-concept study of the BiVAD device, the results are compared to those of the cardiac hand massage. It was also important to see if manually actuating the patches can improve on the results of cardiac hand massage. \Cref{fig:silicone_raw} shows raw data for all three modes of actuation for comparison. For line style, refer to \cref{fig:silicone_raw} \textbf{(d)}. Cardiac hand massage operates at higher flow rates and peak pressures than using the device. The same is true for actuating the patches with hand over using the actuation device. On the other hand, using the device results in lower pressure levels in both ventricles over the other two modes of actuation. 

\begin{figure*}[htbp!]
    \centering
    \includegraphics[scale=1]{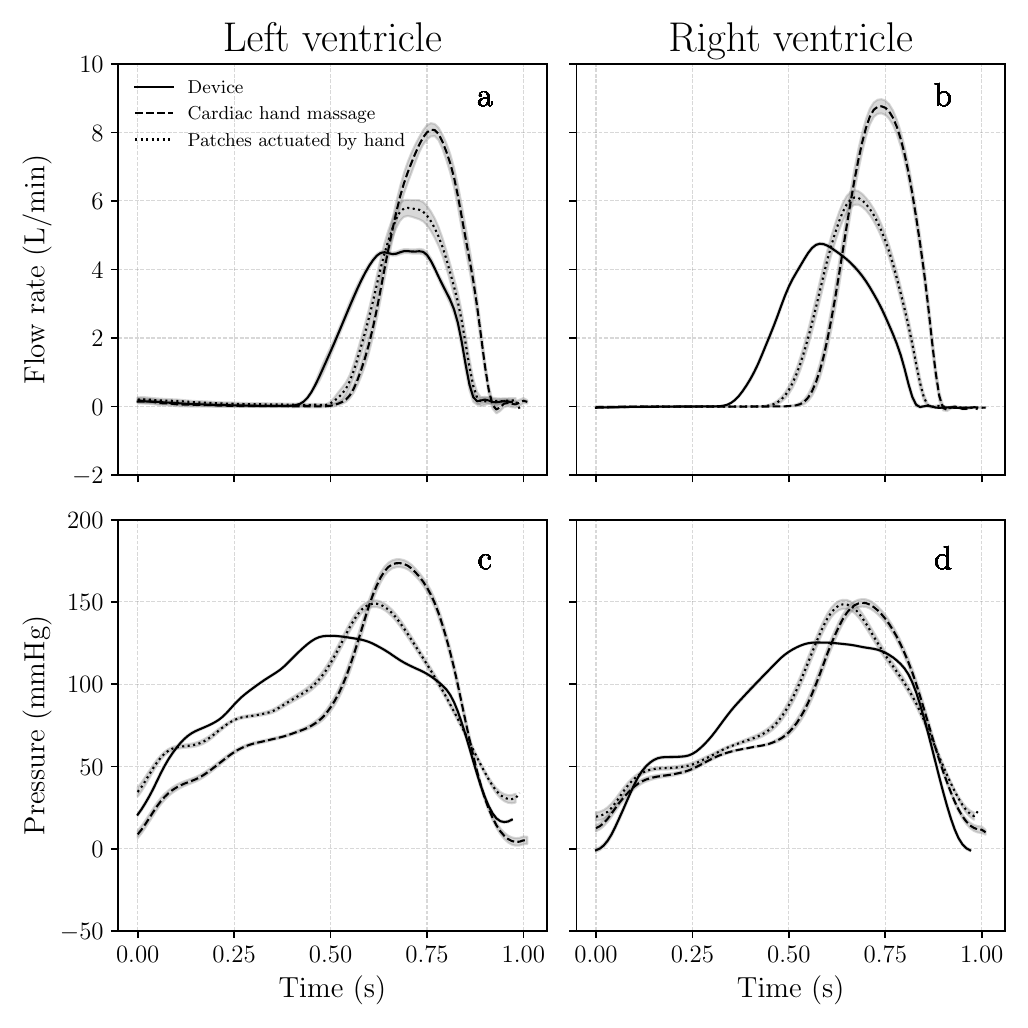}
    \caption{The mean values and the 95\% confidence bands were obtained by sampling individual heart cycles in a measurement. Data was recorded at 60 bpm in all cases. \textbf{a}: mean flow rates in the aorta. \textbf{b}: mean flow rates in the pulmonary artery. \textbf{c}: mean pressures in the left ventricle. \textbf{d}: mean pressures in the right ventricle. The data shows similar trends as \cref{fig:silicone_raw}. Regarding consistency, the device produces the narrowest confidence band.}
    \label{fig:silicone_mean}
\end{figure*}

From \cref{fig:silicone_raw} one cannot draw conclusions on the expected performance and the consistency of the output. This requires statistical studies on the collected data as discussed in Section \ref{postpro}. \Cref{fig:silicone_mean} shows results using means and variation of the data for each mode of actuation. While it shows similar results as \cref{fig:silicone_raw} we now can also see how consistent the different modes are. Regarding consistency, the actuation device produces the least variance among the three modes. 

\begin{figure}[h]
    \centering
    \includegraphics[scale=1]{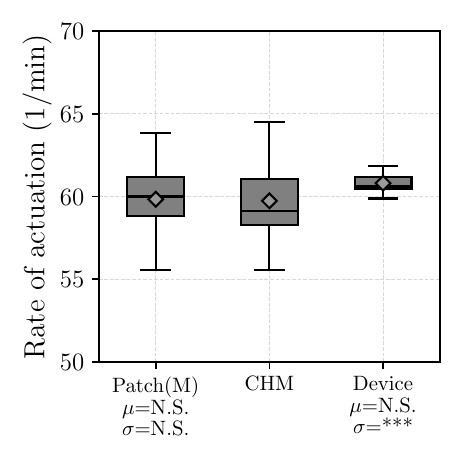}
    \caption{Box plots of the rate of actuation. \textit{Patch(M)} refers to manual actuation of the patches, \textit{CHM} refers to cardiac hand massage, \textit{Device} refers to using the patches with the actuation device (see \cref{fig:modes}). For the statistical symbols, please refer to Table \ref{A_stats}. The means (represented by the rhomboids) do not differ significantly, all three of them are close to the desired 60 bpm value. On the other hand, the variance of the rate of actuation using the device is significantly lower than for the cardiac heart massage. This shows that the actuation using the device is potentially more regular than for the other modes.}
    \label{fig:silicone_box_pulse}
\end{figure}

Essentially, this study compares human operation to machine actuation. Section \ref{postpro} mentions differences in the length of the average cycles between the three modes of actuation. This is the result of disparity in the consistency of actuation. In \cref{fig:silicone_box_pulse} the box plots of the actuation frequency for all three modes are shown. It is visible that the difference between the modes in terms of the means of frequency of actuation is not significant. The variation of the actuation is significantly smaller using the device compared to cardiac heart massage. 

\begin{figure*}[htbp!]
    \centering
    \includegraphics[scale=1]{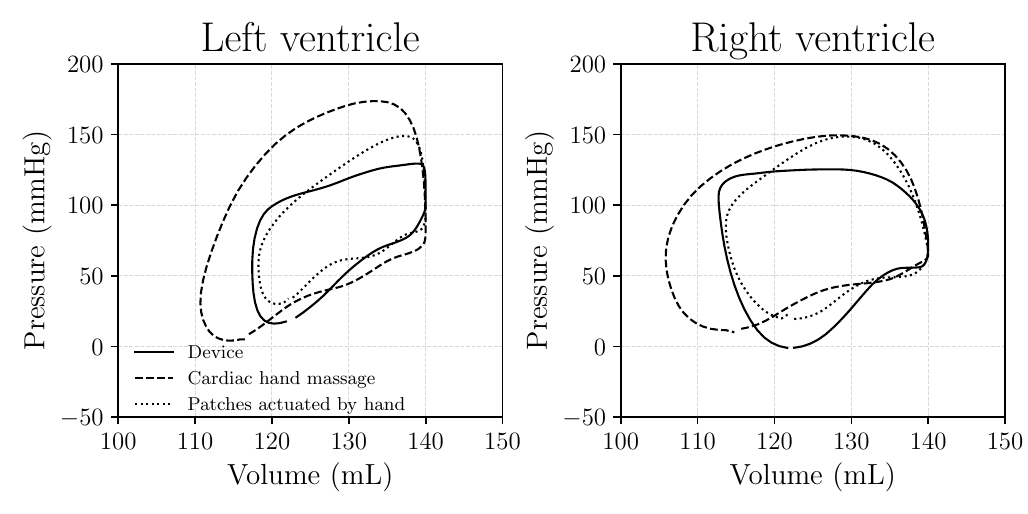}
    \caption{Pressure-volume diagrams of the three actuation modes for both ventricles. The results represent the mean values of pressure and volume based on the entire measurement (see \cref{fig:silicone_mean}). Cardiac hand massage results in higher ejection fraction and higher pressures compared to the modes using the patches. The device provides a similar ejection fraction while producing lower pressures than actuating the patches by hand.}
    \label{fig:silicone_pv}
\end{figure*}

From the averaged values (\cref{fig:silicone_mean}) we can derive the expected pressure-volume diagram for both ventricles. \Cref{fig:silicone_pv} shows these results, from which we notice that the ejected volume is approximately a third of the physiological 95mL \cite{Maceira2006} for both ventricles and all modes. The device produces lower pressure values than the other two modes, while having a similar output as with manual patch actuation. As seen in the previous figures, cardiac hand massage produces higher pressure and higher cardiac output compared to the rest.

\begin{figure*}[htbp!]
    \centering
    \includegraphics[scale=1]{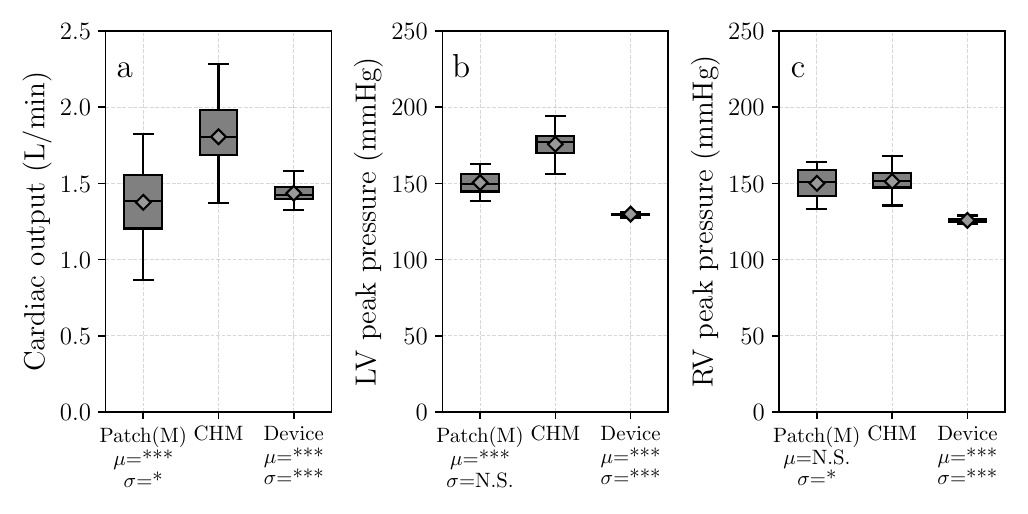}
    \caption{Box plots of the three main performance indicators. \textbf{a}: cardiac output. \textbf{b}: peak pressure in the left ventricle. \textbf{c}: peak pressure in the right ventricle. \textit{Patch(M)} refers to manual actuation of the patches, \textit{CHM} refers to cardiac hand massage, \textit{Device} refers to using the patches with the actuation device (see \cref{fig:modes}). For the statistical symbols, please refer to Table \ref{A_stats}. The expected values (i.e. the mean) are represented by rhomboids. The cardiac output plots show that all three modes can operate around or above the range of 1.5 L/min. The device can keep the peak pressure in the physiological range for the left ventricle, but even though it has significantly lower pressure in the right ventricle than the other two modes, it still greatly exceeds the physiological peak pressures of the right ventricle. From the plots, one can tell that the device has a significantly lower variability over both hand actuation modes in all three measures.}
    \label{fig:silicone_box}
\end{figure*}

For the concluding assessment of the actuation modes, the cardiac output, the peak pressures in the left and right ventricles, and their consistency must be thoroughly examined. \Cref{fig:silicone_box} shows the three properties for all modes of actuation. As seen before, cardiac hand massage produces higher values in all three measures than the modes using the patches. All three modes are close to or exceed the 1.5 L/min cardiac output required to sustain the life of a patient. Using the device, the peak pressures are significantly lower, and all measures show significantly lower variance compared to cardiac hand massage. 

%%==================================%%
%%            Discussion            %%
%%==================================%%

\section{Discussion}\label{discussion}

This proof-of-concept study of a BiVAD system focused on key factors such as cardiac output, pressure values, tissue damage, and the reliability of the BiVAD. All experiments were conducted at a heart rate of 60 bpm to ensure consistency. Results (Section \ref{results}) included data on flow rates and pressures in various cardiac regions, with the heart optimally positioned between the device's patches. Flow rate and pressure data are shown for both ventricles, including raw and post-processed data. Such extensive results are only found in a few publications \cite{Hord2023}. Comparisons were made between three modes of actuation: manual actuation of patches, cardiac hand massage, and using the BiVAD device. The data illustrates higher flow rates and peak pressures with manual methods compared to using the VAD device. However, the device showed lower pressure levels in both ventricles compared to manual methods while maintaining cardiac output close to 1.5 L/min. Overall, the BiVAD system showed promising results with improved consistency over manual actuation modes. Hopefully, our results can contribute to the understanding of how these devices affect the physiology of the cardiovascular system. Further analysis and discussion of these findings are presented in the following section.

Data (\cref{fig:silicone_raw} and \cref{fig:silicone_mean}) indicate that all modes of actuation were relatively consistent. Being machine-operated, consistency is an obvious consequence of using the BiVAD system, but surprisingly, manual operation and cardiac hand massage also show consistent trends. With the latter two modes, we used a metronome to signal 60 bpm pacing for actuation. Prior to the final measurements, we conducted experiments at 40 and 80 bpm, but 60 bpm was chosen based on cardiac output and consistency of the output. Using audio pacing during resuscitative thoracotomy might improve the consistency of the output.

Even though the BiVAD device shows a significantly lower peak flow rate, it produces higher cardiac output compared to manual patch actuation. The device provides flow over a longer period, a feature that can be adjusted by the ratio of $T_{systolic} / T_{diastolic}$ (we used the ratio 2:1).

The BiVAD system managed to keep peak pressure in the LV in the physiological range; however, it failed to do so for the RV. As mentioned in Section \ref{intro}, controlling pressure in both ventricles is crucial for these devices. Other publications usually do not report on the pressure levels in the pulmonary circulation \cite{Letsou2022}, or when they do, the additional pressure is similar for the systemic and pulmonary circulations \cite{Hord2023}. We wanted to be transparent about our findings and limitations regarding the pulmonary circulation (related to the RV).

Something worth noting regarding the ejected volume from the ventricles: \cref{fig:silicone_pv} shows different behavior for the LV and the RV for all modes of actuation. The figure shows a conservative system, excluding implications of measurement errors. As the wall of the LV is thicker than that of the RV, \cref{fig:silicone_pv} shows lower end-systolic volume for the LV. Usually, the RV has a lower stroke volume than the LV \cite{Maceira2006}.

There are indications regarding the quality of recorded data on the box plot diagrams (\cref{fig:silicone_box}). As the mean and median values are almost identical, we can assume negligible skewness in the measured properties. This means that the cardiac output and peak pressures provide similar distributions to normal distribution.

As shown in \cref{fig:silicone_box}, the device was able to operate in the critical 1.5 L/min cardiac output range to sustain the life of a patient. Not only could it produce sufficient cardiac output, but it also managed to achieve this with significantly lower peak pressures and it produced much more consistent heart cycles than cardiac hand massage. This implies that using the BiVAD device provides safer, more consistent but physiologically still sufficient support for patients than open heart massage. Given its unique design, our BiVAD patch system can open new directions in direct mechanical ventricular actuation.

Regarding the safety of operation, our preliminary experiments on porcine heart samples showed that the vacuum cups leave only a temporary mark on the surface of the heart. However, these are results from ex vivo experiments. To further assess the effect of the device on the heart, in vivo experiments are required. 

%%==================================%%
%%   Limitations and Future Work    %%
%%==================================%%

\section{Limitations and Future Work}\label{limit}

We have managed to improve some of the flaws of the previous designs by Kummer et al. These include ballooning of the apex, the transmission cable cutting into the epicardium, and rotation of the patches during compression. The current design is safe, but there is a potential for improvements. As we have found during our experiments on cardiac massage, using our fingertips to compress the LV and at the same time softly compressing the RV with the balls of the hands leads to better flow and pressure distribution between the LV and the RV. Adapting the design of the patch system to mimic this kind of compression can lead to improvements in both cardiac output and peak pressure values. 

Further tests are required using porcine heart samples to test the performance of the device. As the device was showing promising results on silicone hearts, we expect to be able to reproduce these on porcine heart samples. These samples require special preparation to be able to record flow and pressure data on our test bench. So far we have not managed to acquire watertight porcine samples with functioning heart valves. With an improved system, following tests on ex vivo porcine samples, a cadaver experiment should provide new insights into the performance of the BiVAD system.

Biocompatibility and the compactness of the device are not the focus of the research, but will be essential for in vivo tests and future patients.

%%==================================%%
%%            Conclusion            %%
%%==================================%%

\section{Conclusions}\label{conclusions}

In this study, we investigated the capabilities of a novel biventricular assist device (BiVAD) to replace open heart massage. 

The device showcases a unique approach for VADs focusing on resuscitative thoracotomy applications. There exist other devices to support patients via epicardial compression, but our design is unique as it uses mechanical actuation through a patch system. The approach to use these devices in emergency cases rather than as bridge-to-transplant might serve further development of similar designs. The system of the patches, actuation device, and vacuum pump show promising results for a proof-of-concept study.

With a series of experiments, we were able to show promising initial results with our external BiVAD. Our design not only shows a new path for product development, but we also managed to collect data that usually is not published for such devices. Recording pressure and flow rates for both ventricles, and creating reasonable statistics with the recorded data is of interest for the field, yet these results are rarely published. 

The device not only provides cardiac output within the range of 1.5 L/min, but it also achieves this with safer pressure levels and greater consistency compared to traditional cardiac hand massage. Our BiVAD system holds the potential to improve the currently low success rate associated with open heart massage during resuscitative thoracotomy. By supporting essential life functions and relieving medical staff from the burden of manual massage, it allows them to focus on delivering care where it is most needed.

\backmatter

\bmhead{Supplementary information}
Videos of the device can be found in the supplementary information.

\bmhead{Acknowledgments}
The authors thank 
Cardiocentro Ticino for the EmTec BioPro TT clamp-on sensor.
Jan Vincent Spitzer for designing and building the data acquisition system, and assisting with the experiments during his Master's thesis in our institute.
Irma Burazorovic for her research in finding \textit{Dragon Skin \texttrademark} silicone.
Elia Morello for manufacturing the silicone heart samples using \textit{Dragon Skin \texttrademark}.
Shangyi Cao for assisting with experiments.
Johanna Jakab for reviewing the manuscript.
Dr. med. Thomas Theologou for providing prepared porcine hearts for our experiments.
Sebastian Pilz and Dr. med. Giovanni Colacicco from the Anatomical Institute of the University of Zurich for providing Thiel solution to preserve our porcine heart samples. 
Stadt Zürich, Umwelt- und Gesundheitsschutz, Veterinaerdienste for providing porcine hearts.
USZ and LifeTec for earlier cadaver and ex vivo experiments.

\section*{Statements and Declarations}

\bmhead{Author Contributions}
\textbf{Kristóf Sárosi}: new patch and actuation prototype design, development of a new data acquisition system, experimental design, conducting experiments using porcine and silicone hearts, data post-processing, analysis of results, manuscript writing.
\textbf{Thomas Kummer}: provided the base of this research: initial patch and actuation prototypes, initial test bench and data acquisition system, silicon heart design and manufacturing process; advised on experimental procedures, manuscript revision.
\textbf{Thomas Rösgen}: advisory role in data acquisition design, sensor employment and programmed data acquisition elements, manuscript revision.
\textbf{Stijn Vandenberghe}: validation plan review, supported device and experiment design, results evaluation, manuscript revision.
\textbf{Stefanos Demertzis}: validation plan review, results evaluation, defined clinical application limits, manuscript revision.
\textbf{Patrick Jenny}: supervisor, funding, design advice, results evaluation, drafted and revised the manuscript.

\bmhead{Funding}
Open access funding provided by Swiss Federal Institute of Technology Zurich. Internal funding of the Institute of Fluid Dynamics, Swiss Federal Institute of Technology Zurich.

\bmhead{Conflict of interest}
Authors Kristóf Sárosi, Thomas Kummer, Thomas Rösgen, Stijn Vandenberghe, Stefanos Demertzis, and Patrick Jenny declare that they have no conflict of interest. 

\bibliography{sn-bibliography}% common bib file

\clearpage
%%==================================%%
%%            Appendix              %%
%%==================================%%
\begin{appendices}
\onecolumn

\section{Actuation}\label{A_actuation}

\begin{table}[h]
    \centering
    \caption{Parts list of the actuation box, listing major parts.}
    \label{tab:A_actuation}
    
    \begin{tabular}{|l|c|l|}
    \hline
       Function & Qty & Name \\
     \hline
     \hline
        Digital servo motor & 2 &  ANNIMOS DS5160SSG 180° (60kg$\cdot$cm) \\
     \hline
        Servo controller & 1 & Arduino Nano 5V ATmega328P  \\
     \hline
        Main controller & 1 & Arduino Uno WiFi Rev2  \\
     \hline
    \end{tabular}
    
\end{table}

\section{Data Acquisition}\label{A_data}

\begin{table}[h]
    \centering
    \caption{Parts list of the data acquisition system, listing major parts.}
    \label{tab:A_actuation}
    
    \begin{tabular}{|l|c|l|}
    \hline
       Function & Qty & Name \\
     \hline
     \hline
        In-line ultrasonic flow meter & 4 &  Audiowell HS0016 \\
     \hline
        Evaluation board (Audiowell) & 4 & TI EVM430-FR6047  \\
     \hline
        Clamp-on ultrasonic flow meter & 1 & EmTec BioProTT  \\
     \hline
        Evaluation board (EmTec) & 1 & EmTec DIGIFLOW-EXT1  \\
     \hline
        Flexible tubing & 4 & Saint-Gobain Tygon E-3603 3/4" \\
     \hline
        Data processing board & 2 & Arduino Portenta H7  \\
     \hline
        Connectivity board & 2 & Arduino Portenta H7 Breakout Board  \\
     \hline
        Analog pressure sensor & 4 & Honeywell ABPMRNT005PGAA3  \\
     \hline
    \end{tabular}
    
\end{table}

The flow and pressure data within the system are obtained through the use of four digital UART ultrasonic flow meters and four analog pressure sensors, with simultaneous acquisition being a key requirement. 

\section{Statistical Symbols}\label{A_stats}

\begin{table}[h]
\centering
    \caption{Table of statistical symbols}
    \label{table:A_signif}
    \begin{tabular}{ |l | l| }
        \hline
         Symbol & Meaning \\
         \hline
         \hline
         $\mu$ & Probability of accepting the null hypothesis regarding means \\
         $\sigma$ & Probability of accepting the null hypothesis regarding variance \\
         \hline
         N.S. & Not significant \\ 
         * & $p<0.05$ \\  
         ** & $p<0.01$ \\
         *** & $p<0.001$ \\
         \hline
    \end{tabular}
    
\end{table}

\end{appendices}

\end{document}